\documentclass[12pt]{article}
\usepackage{multicol}
\usepackage{makeidx}  % allows for index generation
\usepackage{amssymb}
\usepackage{amsmath}
\usepackage{indentfirst}
\begin{document}
\title{Magnetic resonance in an elliptic magnetic field}
\author{E. A. Ivanchenko \footnote{E-mail: yevgeny@kipt.kharkov.ua}}
\date{National Science Center ``Kharkov Institute of
Physics and Technology'', Institute for Theoretical Physics,\\
Kharkov, Ukraine\\
April 20, 2004}
%\date{Received: date / Revised version: date}
%% The correct dates will be entered by the editor
\maketitle \index{Ivanchenko@Eugene Ivanchenko}

%%%%%%%%%%%%%%%%%%%%%%%%%%%%%%%%%%%%%%%%%%%%%%%%%%%%%%%%%%%%%%%%%%%%%%
\begin{abstract}
The behaviour of a particle  with a spin 1/2 and a dipole magnetic
 moment in a time-varying magnetic field in the  form
 $(h_0 cn(\omega t,k), \\h_0 sn(\omega t,k), H_0 dn(\omega t,k)
)$, where $\omega$ is
 the driving field frequency, $t$ is the time, $h_0$ and $H_0$ are the field
 amplitudes, $cn$, $sn$, $dn$ are Jacobi elliptic functions, $ k$ is the
 modulus of the elliptic functions has been considered. The variation parameter $k$ from
 zero to 1 gives rise to a wide set of functions from trigonometric shapes to exponential pulse shapes  modulating the
 field. The problem was reduced  to the solution of general
 Heun' equation. The exact solution of the wave function was found
 at  resonance for any $ k$. It has been shown that the transition
 probability in this case  does not depend on $k$.
 The present study may be useful for analysis interference experiments,
improving magnetic spectrometers and the field of quantum
computing.
\end{abstract}

%%%%%%%%%%%%%%%%%%%%%%%%%%%%%%%%%%%%%%%%%%%%%%%%%%%%%%%%%%%%%%%%%%%%%%

\section{Introduction}
 Rabi \cite{r1} studied the temporal dynamics of a particle
featuring a dipole magnetic moment $\frac{1}{2}$ in a constant
magnetic field $H_0$, directed along the $z$-axis, and another
varying magnetic field $H_x=h_0cos\omega t,\ H_y=h_0sin\omega t$
rotating with a frequency $\omega$ perpendicular to $H_0$ ( $H_0,\
h_0$ are the amplitudes). There are several methods of modulating
magnetic fields while studying the phenomenon of magnetic
resonance \cite{r2}. This work focuses on the temporal evolution
of a particle with a dipole magnetic moment in a distorted
magnetic field described by $\vec{H}(t)= (h_0 cn(\omega t,k), h_0
sn(\omega t,k), H_0 dn(\omega t,k) )$. Such field modulation under
a changing modulus $k$ of the elliptic functions from zero to
unity describes an entire class of field shapes from trigonometric
\cite{r1} to pulsed
exponential \cite{r3,r4}. \\
 \indent The Schr\"{o}dinger equation of a wave function $\Psi(t)$ that
describes the dynamics of a particle featuring a spin of
$\frac{1}{2}$ and a magnetic moment in a time-varying magnetic
field $\vec H(t)$ is given
\begin{equation}
\label{eq1} i \hbar\partial_t\Psi\left( {t}
\right)=\frac{g\mu_0}{2} \vec{\sigma} \vec{H}({t})\Psi\left( {t}
\right),
\end{equation}
where ${g}$ is the Lande factor, $\mu _{0} $ is the Bohr magneton,
and the Pauli matrices are
\\
$\vec{\sigma}=\left(\sigma_x,\sigma_y,\sigma_z
\right)=\left(\begin{pmatrix}
  _{0} & _{1} \\
  _{1} & _{0}
\end{pmatrix}, \begin{pmatrix}
  _{0} & _{-i} \\
  _{i} & _{0}
\end{pmatrix}, \begin{pmatrix}
  _{1} & _{0} \\
  _{0} & _{-1}
\end{pmatrix}\right)$ . The solution to the Schr\"{o}dinger equation
 could be found by expanding in
 eigenfunctions of matrix  $\sigma_z$
\begin{equation}\label{2}
  \Psi\left( {t}
\right)=\begin{pmatrix}
  _{1} \\
  _{0}
\end{pmatrix}\Psi_1\left( {t}
\right)+\begin{pmatrix}
  _{0} \\
  _{1}
\end{pmatrix}\Psi_2
\left( {t} \right).
\end{equation}\\
Functions $\Psi_1\left( {t} \right)$ and $\Psi_2\left( {t}
\right)$ describe the probability amplitudes without a spin flip
and with one, respectively, and by definition obey the
normalization restriction
\begin{equation}\label{3}
  |\Psi_1\left( {t}
\right)|^2+|\Psi_2\left( {t} \right)|^2=1
\end{equation}
and are assumed to be defined at the initial time $t$.
\\
Considering the condition (2), Eq. (1) takes form
\begin{equation}\label{4}
  i\partial_t\begin{pmatrix}
    _{\Psi_1\left( {t}
\right)} \\
    _{\Psi_2\left( {t}
\right)} \
  \end{pmatrix}=\begin{pmatrix}
    _{Hdn(\omega t,k)} & _{h\left(cn(\omega t,k)-isn(\omega t,k)\right)} \\
    _{h\left(cn(\omega t,k)+isn(\omega t,k)\right)} & _{-Hdn(\omega t,k)
    } \
  \end{pmatrix}\begin{pmatrix}
    _{\Psi_1\left( {t}
\right)} \\
    _{\Psi_2\left( {t}
\right)} \
  \end{pmatrix},
\end{equation}
\begin{equation}\label{5}
H=\frac{g\mu_0H_0}{2\hbar},\  h=\frac{g\mu_0h_0}{2\hbar}.
\end{equation}

\section{Independence of  resonance on modulus $k$}
\indent Let us turn to a dimensionless independent variable
$\tau=\omega t$ and substitute the dependent variables:
\begin{equation}\label{6}
\begin{pmatrix}
    _{\Psi_1\left( {\tau}
\right)} \\
    _{\Psi_2\left( {\tau}
\right)} \
  \end{pmatrix}=\begin{pmatrix}
    _{f} & _{0} \\
    _{0} & _{f^*} \
  \end{pmatrix}\begin{pmatrix}
    _{\varphi_1\left( {\tau}
\right)} \\
    _{\varphi_2\left( {\tau}
\right)} \
  \end{pmatrix},
\end{equation}
 where
\begin{equation}\label{7}
f=\sqrt{cn\tau-isn\tau}=\sqrt{\frac{1+cn\tau}{2}}-isign\left(sn\tau\right)
\sqrt{\frac{1-cn\tau}{2}}.
\end{equation}
The system in (4) takes on a different form through change in (6)
\begin{equation}\label{8}
  i\partial_\tau\begin{pmatrix}
    _{\varphi_1\left( {\tau}
\right)} \\
    _{\varphi_2\left( {\tau}
\right)} \
  \end{pmatrix}_\tau=\begin{pmatrix}
    _{\frac{\Delta}{\omega}dn\tau} & _{\frac{h}{\omega}} \\
    _{\frac{h}{\omega}} & _{-\frac{\Delta}{\omega}dn\tau} \
  \end{pmatrix}\begin{pmatrix}
    _{\varphi_1\left( {\tau}
\right)} \\
    _{\varphi_2\left( {\tau}
\right)} \
  \end{pmatrix},
\end{equation}
in which detuning $\Delta$  is
\begin{equation}\label{9}
  \Delta=H-\frac{\omega}{2}.
\end{equation}
The system in (8) may become similar to one studied by Shirley
\cite{r5} through a time independent orthogonal transformation
$\begin{pmatrix}
  _{\frac{\sqrt{2}}{2}} & _{\frac{\sqrt{2}}{2}}\\
  _{-{\frac{\sqrt{2}}{2}}} & _{\frac{\sqrt{2}}{2}}
\end{pmatrix}$ .\\
\indent   Let us note the mechanical-geometrical analog of the
system in (8) while presenting the unknown functions as
$\varphi_1(\tau)=x+iy,\ \varphi_2(\tau)=u+iv $, where $x,\ y,\ u,\
v$ are real functions that are defined at the initial time and
satisfy a non-linear system of differential equations with
constant coefficients:
\begin{equation}\label{10}
  x^2+y^2+u^2+v^2=1,
\end{equation}
\begin{equation}\label{11}
  v \partial_\tau x-u \partial_\tau y+y \partial_\tau u -
x \partial_\tau v=\frac{h}{\omega},
\end{equation}
\begin{equation}\label{12}
(\partial_\tau x)^2+(\partial_\tau y)^2+(\partial_\tau
u)^2+(\partial_\tau v)^2+\frac{\Delta^2k^2}{\omega^2}sn^2(\tau,\ k
)=\frac{h^2+\Delta^2}{\omega^2},
\end{equation}
\begin{equation}\label{13}
y \partial_\tau x-x \partial_\tau y+u \partial_\tau v -
 v \partial_\tau u=\frac{\Delta}{\omega}dn(\tau,\ k),
\end{equation}
 which describe the motion of a 4-vector $(x, y, u, v)$ along
 the spherical surface (10) under two conservation restrictions (11), (12).\\
 \indent The equation which is required to determine function $\varphi_2(\tau)$ could be
found from the system (8)
 \begin{equation}\label{14}
  \partial_{\tau\tau}\varphi_2(\tau)+\left(i\frac{\Delta}{\omega}k^2{sn\tau}{án\tau}-\frac{\Delta^2}{\omega^2}k^2{sn^2\tau}+\frac{\Omega^2_R}{\omega^2}
\right)\varphi_2(\tau)=0,
\end{equation}
where
\begin{equation}\label{15}
  \Omega^2_R=h^2+\Delta^2,
\end{equation}
and is presented as a generalized Lame equation in the Jacobi form.\\
Eq. (14) shows that the real  $u$ and the imaginary $v$ components
of function $\varphi_2(\tau)$ define a gyroscopic system with a
parametric excitation of its intrinsic frequency
$\frac{h}{\omega}$ \cite{r6}. \\
\indent Let $f_1(\tau)$ and $f_2(\tau)$  become a fundamental
system that defines a general solution to Eq. (14)
\begin{equation}\label{16}
 \varphi_2(\tau)=Af_1(\tau)+Bf_2(\tau),
\end{equation}
where $A$, $B$ are arbitrary constants. The solution of the Cauchy
problem (1) could be expressed through functions $f_1(\tau)$ and
$f_2(\tau)$ using the initial conditions $\begin{pmatrix}
  _{\Psi_1(0)} \\
  _{\Psi_2(0)}
\end{pmatrix}$ as
\begin{equation}\label{17}
\begin{pmatrix}
  _{\Psi_1(\tau)} \\
  _{\Psi_2(\tau)}
\end{pmatrix}=\begin{pmatrix}
  _{f} & _{0} \\
  _{0} & _{f^*}
\end{pmatrix}\begin{pmatrix}
  _{a_1F_1+b_1F_2} & _{a_2F_1+b_2F_2} \\
  _{a_1f_1+b_1f_2} & _{a_2f_1+b_2f_2}
\end{pmatrix}\begin{pmatrix}
  _{\Psi_1(0)} \\
  _{\Psi_2(0)}
\end{pmatrix},
\end{equation}
where
\begin{equation}\label{18}
\begin{split}
  F_{1,2}=\frac{i\partial_\tau{f_{1,2}}+\frac{\Delta}{\omega}f_{1,2}dn\tau}{\frac{h}{\omega}},a_1=\frac{f_2(0)}{d},a_2=-\frac{F_2}{d}|_{\tau=0},\\
b_1=-\frac{f_1(0)}{d},b_2=\frac{F_1}{d}|_{\tau=0},
\frac{h}{\omega}d=-i(f_1\partial_\tau{f_2}-f_2\partial_\tau{f_1})|_{\tau=0}.
\end{split}
\end{equation}
Assuming that the wave function $\begin{pmatrix}
  _{\Psi_1(0)} \\
  _{\Psi_2(0)}
\end{pmatrix}$ is equal to $\begin{pmatrix}
  _{1} \\
  _{0}
\end{pmatrix}$ at the initial
time, the solution for the wave function (17) at time $\tau$ is
given by
\begin{equation}\label{19}
  \begin{pmatrix}
    _{\Psi_1\left( {\tau}
\right)} \\
    _{\Psi_2\left( {\tau}
\right)} \
  \end{pmatrix}=\begin{pmatrix}
    _{f} & _{0} \\
    _{0} & _{f^{*}} \
  \end{pmatrix}\begin{pmatrix}
    _{a_1F_1+b_1F_2} \\
    _{a_1f_1+b_1f_2} \
  \end{pmatrix}.
\end{equation}
The probability of a transition requiring a spin flip over time
$\tau$ is
    \begin{equation}\label{20}
  P_{\frac{1}{2}\rightarrow-\frac{1}{2}}(\tau,\Delta,k)=|a_1f_1+b_1f_2|^2.
\end{equation}
Thus, the formula describing the transition probability could be
expressed through functions $f_1(\tau),f_2(\tau)$ using the
formulae in (18) and (20) as
\begin{equation}\label{21}
  P_{\frac{1}{2}\rightarrow-\frac{1}{2}}(\tau,\Delta,k)=\frac{h^2}{\omega^2}\left|\frac{f_1(\tau)f_2(0)-f_2(\tau)f_1(0)}
{(f_1(\tau)\partial_\tau{f_2(\tau)}-f_2(\tau)\partial_\tau{f_1(\tau)})|_{\tau=0}}\right|^2.
\end{equation}
The Rabi result \cite{r1} provided by Eq. (14) at $k=0$ stipulates
that $f_1(\tau)=\cos\frac{\Omega_R}{\omega}\tau,\
f_2(\tau)=\sin\frac{\Omega_R}{\omega}\tau$, and the transition
probability is
\begin{equation}\label{22}
P_{\frac{1}{2}\rightarrow-\frac{1}{2}}(\tau,\Delta,k=0)=
\frac{h^2}{\Omega_R^2}\sin^2{\frac{\Omega_R}{\omega}}\tau.
\end{equation} \\
Eq.(14) is simplified at $0\leq{k}\leq1$ in case of a sharp
fundamental resonance with $\Delta=0$:
 \begin{equation}\label{23}
  \partial_{\tau\tau}\varphi_2(\tau)+\frac{h^2}{\omega^2}\varphi_2(\tau)=0.
\end{equation}
Therefore, $f_1(\tau,\Delta=0,k)=\cos\frac{h}{\omega}\tau,\
f_2(\tau,\Delta=0,k)=\sin\frac{h}{\omega}\tau$ and Eq. (19) is
solved explicitly
\begin{equation}\label{24}
\begin{pmatrix}
    _{\Psi_1\left( {\tau}
\right)} \\
    _{\Psi_2\left( {\tau}
\right)} \
  \end{pmatrix}=\begin{pmatrix}
    _{f \cos\frac{h}{\omega} \tau} \\
    _{-if^* \sin\frac{h}{\omega} \tau} \
  \end{pmatrix}.
\end{equation}
Obviously, the transition probability is independent of the $k$
modulus and given by
\begin{equation}\label{25}
P_{\frac{1}{2}\rightarrow-\frac{1}{2}}(\tau,\Delta=0,0\leq{k}\leq1)=
sin^2\frac{h}{\omega}\tau.
\end{equation}
\indent \textit{The fundamental resonance is stable at any value
of the $k$ modulus with respect to consistent variations of the
longitudinal and traversal magnetic fields. In other words, a
distortion in the traversal field is fully compensated
by a corresponding distortion in the longitudinal field}  .\\
\indent Knowing the wave function in (24) makes it possible to
find the polarization vector  $\vec{P}$ defined by
\begin{equation}\label{26}
  P_i=\left( \Psi(t) {\sigma}_i \Psi(t) \right) \quad (i=x,\ y,\
z).
\end{equation}
A simple calculation produces
\begin{equation}\label{27}
  \vec {P}=
({sn{\gamma_m H_0 t}}sin{\gamma_m h_0 t },-{cn{\gamma_m H_0
t}}sin{\gamma_m h_0 t },cos{\gamma_m h_0 t }).
\end{equation}
The polarization vector satisfies the Bloch equation
\begin{equation}\label{28}
  \partial_t\vec{P}=\gamma_m[\vec{H},
\vec{P}],\quad\gamma_m=\frac{g\mu_0}{\hbar}.
\end{equation}

\section{Reduction to Heun equation}
 A general case where both the detuning (9) and the modulus of
the elliptical functions are different from zero requires changing
the variable in Eq. (14) and taking advantage of the doubly
periodicity of
the elliptical functions.\\
It is known that
\begin{equation}\label{29}
 sn(\tau^{'}+iK^{'})=\frac{1}{ksn\tau^{'}},\quad
cn(\tau^{'}+iK^{'})=-i\frac{dn\tau^{'}}{ksn\tau^{'}},\quad
0<{k}<1,
\end{equation}
where the full elliptical integral of the first kind $K^{'}$ is
\begin{equation}\label{30}
  K^{'}=\int_{0}^{\frac{\pi}{2}}\frac{1}{\sqrt{1-{k^{'}}^2sin^2\varphi}}\,d\varphi
,\quad k^2+k{'}^2=1.
\end{equation}
Let us switch to a half-variable in functions $sn\tau^{'},\
dn\tau^{'}$ using
\begin{equation}\label{31}
  sn\tau^{'}=\frac{2sn\frac{\tau^{'}}{2}cn\frac{\tau^{'}}{2}dn\frac{\tau^{'}}{2}}
{1-k^2sn^4\frac{\tau^{'}}{2}},\quad
dn\tau^{'}=\frac{dn^2\frac{\tau^{'}}{2}-k^2sn^2\frac{\tau^{'}}{2}cn^2\frac{\tau^{'}}{2}}
{1-k^2sn^4\frac{\tau^{'}}{2}}
\end{equation}
and introducing a new independent variable
\begin{equation}\label{32}
  z=sn^2\frac{\tau^{'}}{2}=k^{-\frac{1}{2}}\frac{(1+k)sn\frac{\tau}{2}-
icn\frac{\tau}{2}dn\frac{\tau}{2}}{1+ksn^2\frac{\tau}{2}},
\end{equation}
we arrive at the algebraic form of an equation for function
$\varphi_2(\tau) \equiv y(z)$ after simple transformations:
\begin{equation}\label{33}
\begin{split}
  \partial_{zz}y+
\left(\frac{A_1}{z}+\frac{A_2}{z-1}+\frac{A_3}{z-\frac{1}{k^2}} \right)\partial_{z}y+\\
\left(\frac{B_1}{z^2}+\frac{B_2}{(z-1)^2}+\frac{B_3}{(z-\frac{1}{k^2})^2}+\frac{C_1}{z}+\frac{C_2}{z-1}+\frac{C_3}{z-\frac{1}{k^2}}
\right)y=0,
\end{split}
\end{equation}
where
\begin{equation}\label{34}
\begin{split}
  A_1=A_2=A_3=\frac{1}{2},\, B_1=B_2=a-b,\, B_3=-(a+b),\, \\
a=\frac{\Delta}{4\omega},\, b=\frac{\Delta^2}{4\omega^2}.
\end{split}
\end{equation}
\begin{equation}\label{35}
  C_1=2(a-b)-2bk^2+\frac{\Omega^2_R}{\omega^2},
\end{equation}
\begin{equation}\label{36}
  C_2=-2(a-b)-(a+b)+\frac{1}{k^2-1}[a(k^2-1)-b(k^2+1)+\frac{\Omega^2_R}{\omega^2}],
\end{equation}
\begin{equation}\label{37}
 C_3=k^2(a+b)-\frac{1}{k^2-1}[a(k^2-1)-b(k^2+1)+\frac{\Omega^2_R}{\omega^2}].
\end{equation}
The values of $C_1,\ C_2,\ C_3 $ are linked through
\begin{equation}\label{38}
  C_1+C_2+C_3=0.
\end{equation}
The functional form of the coefficients in Eq. (33) under the
condition in (38) is necessary and sufficient to classify this
equation as Fuchs'  \cite{r7}.\\ Eq. (33) may be simplified
through substituting the dependent variable \cite{r7,r9}
\begin{equation}\label{39}
  y(z)=w(z)v(z),\  w(z)=z^p(z-1)^q(z-\frac{1}{k^2})^r,
\end{equation}
where  $p,q,r$  are both solutions to the indicial equations and
characteristic exponents in the vicinity of the following points
\begin{equation}\label{40}
 z=\alpha_1=0,\ z=\alpha_2=1,\ z=\alpha_3=\frac{1}{k^2}
\end{equation}
The solutions may be found from the indicial equations:
\begin{equation}\label{41}
  p(p-1)+\frac{1}{2}p+B_1=0 \Rightarrow
p_\pm=\frac{1}{4}\pm|\frac{\Delta}{2\omega}-\frac{1}{4}|,
\end{equation}
\begin{equation}\label{42}
  q(q-1)+\frac{1}{2}q+B_2=0 \Rightarrow
q_\pm=\frac{1}{4}\pm|\frac{\Delta}{2\omega}-\frac{1}{4}|,
\end{equation}
\begin{equation}\label{43}
  r(r-1)+\frac{1}{2}r+B_3=0 \Rightarrow
r_\pm=\frac{1}{4}\pm|\frac{\Delta}{2\omega}+\frac{1}{4}|.
\end{equation}\\
The characteristic exponents in the vicinity of $z=\infty$ are
determined from the indicial equation \cite{r7}\\
\begin{equation}\label{44}
  \rho^{\infty}(\rho^{\infty}-1)+(2-\sum_{k=1}^{3}A_k)\rho^{\infty}+
\sum_{k=1}^{3}(B_k+\alpha_kC_k)=0,
\end{equation}
which takes the following form using (34-37), (40)
\begin{equation}\label{45}
\rho^{\infty}(\rho^{\infty}-1)+\frac{1}{2}\rho^{\infty}+B_3=0
\Rightarrow
\rho^{\infty}_{\pm}=\frac{1}{4}\pm|\frac{\Delta}{2\omega}+\frac{1}{4}|.
\end{equation}
Thus, we obtain the Heun equation with real parameters \cite{r8}
for function $v(z)$ in the Klein-B\^{o}cher-Ince form
\cite{r9,r10}
\begin{equation}\label{46}
\partial_{zz}v+
 \left(
\frac{\gamma}{z}+\frac{\delta}{z-1}+\frac{\epsilon}{z-\frac{1}{k^2}}
\right)\partial_{z}v+ \frac{\alpha \beta z
-q_a}{z(z-1)(z-\frac{1}{k^2})}v=0,
\end{equation}
where
\begin{equation}\label{47}
  \begin{split}
\gamma=2p+\frac{1}{2},\  \delta=2q+\frac{1}{2},\
\epsilon=2r+\frac{1}{2}, \\
\alpha=\rho^{\infty}_{+}+p+q+r,\  \beta=\rho^{\infty}_{-}+p+q+r.
\end{split}
\end{equation}
The accessory parameter $q_a$ is given by
\begin{equation}\label{48}
  q_a=\gamma r+\frac{1}{2}p-\frac{C_1-\gamma q-\frac{1}{2}p}{k^2}.
\end{equation}
It is easy to verify that the Fuchs' condition is satisfied:
\begin{equation}\label{49}
  \gamma+\delta+\epsilon=\alpha+\beta+1.
\end{equation}
The Riemann symbol  that characterizes the form of the solutions
to Heun equation (46) is given by
\begin{equation}\label{50}
P= \begin{pmatrix}
  _{0} & _{1} & _{\frac{1}{k^2}} & _{\infty}  \\
  _{0} & _{0} & _{0} & _{\alpha} & _{z;} & _{q_a} \\
  _{1-\gamma} & _{1-\delta} & _{1-\epsilon} & _{\beta}
\end{pmatrix}.
\end{equation}\\
It is worth noting that the substitution in (39) could be done
eight ways
\begin{equation}\label{51}
 \begin{split}
  pqr\in \{p_+q_+r_+,\ p_+q_-r_+,\ p_+q_+r_-,\ p_+q_-r_-,\\
 p_-q_+r_+,\ p_-q_-r_+,\ p_-q_+r_-,\ p_-q_-r_- \}\,  ,
\end{split}
\end{equation}\\
which sometimes allows to pick suitable characteristic exponents
and accessory parameter  for specific calculations. Thus, the
linearly independent solutions to Eq. (14) are written as $f_1=w
v_1,\ f_2=w v_2 $, where  $v_1,\ v_2$ belong to the fundamental
system of solutions to the Heun equation (46). The probability of
the transition with a spin flip is expressed through the solutions
$v_1,\ v_2$ as :
\begin{equation}\label{52}
  P_{\frac{1}{2}\rightarrow-\frac{1}{2}}(\tau,\Delta,k)=\frac{h^2}{\omega^2}\left|\frac{w(\tau)(v_1(\tau)v_2(0)-v_2(\tau)v_1(0))}
{[w(\tau)(v_1(\tau)\partial_\tau{v_2(\tau)}-v_2(\tau)\partial_\tau{v_1(\tau)})]|_{\tau=0}}\right|^2.
\end{equation}
\hspace*{\parindent} In practice, solutions to the Heun equation
are obtained through a reduction of the parameter space for
$\gamma,\ \delta,\ \epsilon,\ \alpha,\ \beta,\ q_a $ taking into
account that there are additional relationships between the
parameters [9] besides the condition in (49). An analysis of
parametric resonances induced by a non-zero detuning and
distortion of a magnetic field will be presented elsewhere.
\section{Spin $ \geq { \frac{1}{2}}$}
Generalization for higher spin values is not difficult because the
matrix in (17) is unitary due to the hermitian property of the
Hamiltonian in Eq. (1):
 \begin{equation}\label{53}
\begin{pmatrix}
  _{f(a_1F_1+b_1F_2)} & _{f(a_2F_1+b_2F_2)} \\
  _{f^{*}(a_1f_1+b_1f_2)} & _{f^{*}(a_2f_1+b_2f_2)}
\end{pmatrix}
=D^\frac{1}{2}(\varphi,\ \theta,\ \psi),
\end{equation}
where $D^\frac{1}{2}(\varphi,\ \theta,\ \psi)$ is a Wigner matrix.
The Euler angles $\varphi,\ \theta,\ \psi$ are determined from
equations
\begin{equation}\label{54}
f(a_1F_1+b_1F_2)=\cos\frac{\theta}{2}\exp{\frac{i}{2}(\varphi+\psi)},
\end{equation}
\begin{equation}\label{55}
f^{*}(a_1f_1+b_1f_2)=i
\sin\frac{\theta}{2}\exp{\frac{-i}{2}(\varphi-\psi)},
\end{equation}
\begin{equation}\label{56}
  sin^{2}\frac{\theta}{2}=P_{\frac{1}{2} \rightarrow
-\frac{1}{2}}(\tau,\Delta,k)
\end{equation}
and are independent of the magnitude of the angular momentum $J$
\cite{r11}. Therefore, the transition probability over time $\tau$
from the state with a projection of angular momentum $m$ into one
with a projection $m{'}$ for a particle with a spin $J$ is given
by
\begin{equation}\label{56}
  \begin{split}
P_{m \rightarrow m{'}}(\tau,\Delta,k)=|D^{(J)}_{mm{'}}(\varphi,\
\theta,\ \psi)|^2=[(J+m)!(J-m)!(J+m{'})!(J-m{'})!]\\
cos^{4J}\frac{\theta}{2}\left[ \sum_{\nu} (-1)^{\nu}
\frac{({\tan\frac{\theta}{2}})^{2 \nu-m+m{'}}}{ \nu ! (
\nu-m+m{'}) ! (J+m-\nu) ! (J-m{'}-\nu) !} \right]^2.
\end{split}
\end{equation}
The Wigner matrix in formula (57) is
\begin{equation}\label{58}
\begin{split}
D^{(J)}_{mm{'}}(\varphi,\ \theta,\
\psi)=i^{m'-m}e^{i(m\varphi+m'\psi)}{[(J+m)!(J-m)!(J+m{'})!(J-m{'})!]}^ \frac{1}{2}\\
\sum_{\nu} (-1)^{\nu}\frac{({\sin\frac{\theta}{2}})^{2
\nu-m+m{'}}({\cos\frac{\theta}{2}})^{2J-2 \nu+m-m{'}}}{ \nu ! (
\nu-m+m{'}) ! (J+m-\nu) ! (J-m{'}-\nu) !},
\end{split}
\end{equation}
\\
and  $m,\ m'=(-J,\ -J+1,\ ...,\ J-1,\ J)$.\\
\section{Conclusion}
The study presented in this work may be useful in analyzing the
results of interference experiments \cite{r12}, improving magnetic
spectrometers, and the field of quantum computing.

\end{document}